\documentclass[prl,letterpaper,aps,floatfix,twocolumn]{revtex4-1}
\usepackage{graphicx}
\usepackage{amsmath}
\usepackage{subfigure}
\newcommand{\beq}{\begin{equation}}
\newcommand{\eeq}{\end{equation}}
\newcommand{\bk}{{{\bf{k}}}}

\newcommand{\br}{{{\bf{r}}}}

\newcommand{\bA}{{\bf{A}}}

\newcommand{\bq}{{\bf{q}}}

\newcommand{\beqa}{\begin{eqnarray}}
\newcommand{\eeqa}{\end{eqnarray}}

\newcommand{\bnabla}{{\boldsymbol \nabla}} 

\newcommand{\btau}{{\boldsymbol \tau}}
\newcommand{\bOmega}{{\boldsymbol \Omega}}

\begin{document}
\title{Anomalous Hall Effect in Weyl Metals}
\author{A.A. Burkov}
\affiliation{Department of Physics and Astronomy, University of Waterloo, Waterloo, Ontario 
N2L 3G1, Canada} 
\date{\today}
\begin{abstract}
We present a theory of the anomalous Hall effect (AHE) in a doped Weyl semimetal, or Weyl metal, 
including both intrinsic and extrinsic (impurity scattering) contributions. We demonstrate that a Weyl metal is distinguished from 
an ordinary ferromagnetic metal by the absence of the extrinsic and the Fermi surface part of the intrinsic 
contributions to the AHE, as long as the Fermi energy is sufficiently close to the Weyl nodes. The AHE in a Weyl metal is thus shown to be a purely 
intrinsic, universal property, fully determined by the location of the Weyl nodes in the first Brillouin zone. 
\end{abstract}
\maketitle
An exciting recent development in condensed matter physics is the emerging extension of the concepts 
of nontrivial electronic structure topology, which have long been confined exclusively to insulators, 
to gapless {\em metallic} states. 
These ideas, pioneered some time ago by Volovik~\cite{Volovik}, have recently been brought to the
forefront of condensed matter research, with 
specific solid-state realizations of the first topologically-nontrivial metallic state, a {\em Weyl semimetal}, 
proposed~\cite{Wan11,Ran11,Burkov11,Xu11}. 
The recent observation of the 
closely related {\em Dirac semimetals}~\cite{Kane12,Fang12,Fang13,Cava13,Shen13,Hasan13} paves the way
for the realization of Weyl semimetals in the near future.

The electronic structure of a Weyl semimetal contains points in momentum space, at which two nondegenerate 
bands touch at the Fermi energy.
Such points, called Weyl nodes, can occur generically (but not necessarily at the Fermi energy) in three-dimensional (3D) band structures, as long 
as either time-reversal (TR) or inversion (I) symmetries are violated, which is needed to create nondegenerate 
bands, otherwise prohibited by the Kramers theorem. 
These points are topologically-nontrivial objects, characterized by an integer topological charge, 
and are monopole sources of the Berry curvature, momentum-space dual of the magnetic field in real space. 

Apart from the appearance of the Weyl nodes themselves, which is generically possible in 3D, Weyl {\em semimetal} requires
the Fermi energy to be aligned with the nodes. This situation is not generic, but is a special case of a Weyl {\em metal}: 
a metal is which the Fermi surface is broken up into disjoint pieces, each surrounding, in the simplest case, a single Weyl node (we will 
assume such individual sheets of the Fermi surface may be characterized by a Chern number, which is equal to the topological charge, enclosed 
by the Fermi surface sheet). 
One may then ask the following question: what are, if any, observable consequences of such a topologically-nontrivial character of the Fermi 
surface of a Weyl metal? 

The purpose of this paper is to describe one such phenomenon, which is characteristic of a specific subclass of Weyl metals,
namely the ferromagnetic (FM) Weyl metals, in which the nodes owe their existence to broken TR~\cite{footnote3}. 
Any FM metal exhibits anomalous Hall effect (AHE), i.e. an antisymmetric contribution to the off-diagonal resistivity, ``proportional" to the 
magnetization rather than to the applied magnetic field.
As has been clearly demonstrated in recent work~\cite{AHE1,AHE2}, geometrical properties of the electronic structure play an important
role in this effect. 

Perhaps the most controversial part of the AHE story, that has emerged in recent years, has been the relative role played by the intrinsic properties of the electronic structure 
of the material (intrinsic AHE) and impurity scattering (extrinsic AHE). 
In general, both are present and are of the same order of magnitude, making it very difficult to disentangle intrinsic and extrinsic contributions experimentally. 
In this paper we show that, in contrast to a generic FM metal, in a FM Weyl metal the extrinsic contribution is essentially absent and the AHE is of purely 
intrinsic origin. Moreover, the intrinsic part of the AHE is fully determined only by the relative location and topological charge of the Weyl nodes, and is 
(almost) independent of the properties of the Fermi surface, as long as individual Fermi surface sheets have nonzero Chern numbers. 
This property becomes increasingly more precise as the Fermi energy approaches the Weyl nodes. As we show below, this 
is closely related to the topology of the Weyl nodes. 

We start from a simple model of a ferromagnetic Weyl metal, motivated by the topological insulator multilayer model, introduced by us before~\cite{Burkov11}. 
The model has the advantage of being general enough to capture all the essential features of the electronic structure of a generic metallic FM, 
yet simple enough to be amenable to analytic calculations. 
The momentum-space Hamiltonian we start from is given by
\beq
\label{eq:1}
H_t(\bk ) =  v_F (\hat z \times \btau) \cdot \bk + m_t(k_z) \tau^z.
\eeq
Here $\btau$ are Pauli matrices, $t = \pm$, $m_{\pm} = b \pm \Delta(k_z)$, $b$ is the mean-field spin splitting and $\Delta(k_z)$ is the band dispersion along the 
magnetization direction in the first Brillouin zone (BZ) $-\pi/ d \le k_z < \pi/d$, and we will use $\hbar = c = 1$ units throughout. The specific form of $\Delta(k_z)$ is unimportant, but we will assume that it is a nonnegative function with a single minimum and a single maximum in the first BZ at $k_z = 0$ and $k_z = \pi/d$. This guarantees the simplest situation 
with a single pair of Weyl nodes. 
Formally, this may be thought of as a Hamiltonian of a pair of 2D Dirac fermions, with the masses $m_{\pm}(k_z)$, which depend on 
parameter $k_z$~\cite{footnote1}. 
In the paramagnetic state, when $b = 0$, Eq.~\eqref{eq:1} describes two pairs of Kramers-degenerate bands with dispersions 
$\pm \sqrt{v_F^2 (k_x^2 + k_y^2) + \Delta^2(k_z)}$. 
When $b$ becomes sufficiently large (taking $b > 0$ for concreteness), the mass $m_-(k_z)$ may change sign at a minimum of two 
points in the BZ, given by the solutions of the equation $\Delta(k_z) = b$. These points are the Weyl nodes. 
Eq.~\eqref{eq:1} may be regarded as a minimal model of the electronic structure of a 3D metallic FM. 

We would like to evaluate the anomalous Hall conductivity of this model FM, in the presence of impurity potential $V(\br) = V_0 \sum_a \delta(\br - \br_a)$, which we will 
assume for simplicity to be gaussian, with only second order correlators present: $\langle V(\br) V(\br ') \rangle = \gamma^2 \delta(\br - \br ')$, where $\gamma^2 = n_i V_0^2$ and 
$n_i$ is the impurity density.  
The higher-order correlators, which are known to be important for AHE (skew-scattering) in principle, do not in fact affect our results.
We will also assume that the impurity potential is diagonal in both the pseudospin $\btau$ and the $t = \pm$ index. Again, this assumption 
is used only for computational simplicity and does not affect the essence of our results. 
To find the anomalous Hall conductivity, we will use a somewhat nonstandard method, which we find to be the most convenient one for our 
purposes, as it allows to most clearly separate physically distinct contributions to the AHE. 
Namely, we imagine coupling electromagnetic field to the electrons and integrating the electron degrees of freedom out to obtain an 
effective action for the electromagnetic field, which, at quadratic order, describes the linear response of the system. 
The part of this action we are interested in has the appearance of a Chern-Simons term, which, adopting the Coulomb gauge for the electromagnetic vector potential $\bnabla \cdot \bA = 0$, is given by
\beq
\label{eq:2}
S = \sum_{\bq, i\Omega}\epsilon^{z 0 \alpha \beta} \Pi(\bq, i\Omega) A_0(-\bq, -i\Omega) \hat q_{\alpha} A_{\beta}(\bq, i \Omega),
\eeq 
where $\hat q_{\alpha} = q_{\alpha}/ q$ and summation over repeated indices is implicit. The $z$-direction in Eq.~\eqref{eq:2} is picked out by the 
magnetization $b$. As we will be interested specifically in the zero frequency and zero wavevector limits of the response function $\Pi(\bq, i \Omega)$, 
we will assume henceforth that $\bq = q \hat x$, which does not lead to any loss of generality due to full rotational symmetry in the $xy$-plane. 
The anomalous Hall conductivity is given by the zero frequency and zero wave vector limit of the response function $\Pi(\bq, i \Omega)$ as
\beq
\label{eq:3}
\sigma_{xy} = \lim_{i \Omega \rightarrow 0} \lim_{q \rightarrow 0} \frac{1}{q} \Pi(\bq, i \Omega).
\eeq
An advantage of Eq.~\eqref{eq:3}, compared to the more standard Kubo formula for the anomalous Hall conductivity, which relates 
it to the current-current correlation function, is that Eq.~\eqref{eq:3} ties the Hall conductivity to the response of a conserved quantity, 
i.e. the particle density. 

The impurity average of the response function $\Pi(\bq, i\Omega)$ may be evaluated by the standard methods of diagrammatic perturbation theory. 
Due to our assumption that the impurity potential is diagonal in the band index $t$, we can do this calculation separately for each pair of bands, labeled 
by $t$, and then simply sum the individual contributions. We will thus omit the $t$ index in what follows, until we come to the final results. 
The retarded impurity averaged one-particle Green's functions have the following general form
\beq
\label{eq:4}
G^R_{\sigma_1 \sigma_2}(\bk, \epsilon) = \frac{z^s_{\bk \sigma_1} \bar z^s_{\bk \sigma_2}}{\epsilon - \xi^s_\bk + i /2 \tau_s}. 
\eeq
Here $s = \pm$ labels the two bands, obtained by diagonalizing Eq.~\eqref{eq:1} for a specific $t$ (the sum over $s$ is made implicit above), $\xi^s_\bk = s \epsilon_\bk - \epsilon_F = s \sqrt{v_F^2(k_x^2 + k_y^2) + m^2(k_z)} - \epsilon_F$ are the band energies, counted from the Fermi energy $\epsilon_F$, and $| z^s_\bk \rangle = \frac{1}{\sqrt{2}} \left(\sqrt{1 + s \frac{m(k_z)}{\epsilon_\bk}}, - i s e^{i \varphi} \sqrt{1 - s \frac{m(k_z)}{\epsilon_\bk}}\right)$
is the corresponding eigenvector with $e^{i \varphi} = \frac{k_x + i k_y}{\sqrt{k_x^2 + k_y^2}}$. 
In what follows we will assume, for concreteness, that $\epsilon_F > 0$, i.e. the Weyl metal is electron-doped (all the results are independent of the sign of the Fermi energy due to 
particle-hole symmetry of our model). 
The impurity scattering rates $1/\tau_{\pm}$ are given, in the Born approximation, by 
\beq
\label{eq:6}
\frac{1}{\tau_s(k_z)} = \frac{1}{\tau}  \left[1 +  s \frac{m(k_z) \langle m \rangle}{\epsilon_F^2} \right],
\eeq
where $1/\tau = \pi \gamma^2 g(\epsilon_F)$ and $g(\epsilon_F) = \int \frac{d^3 k}{(2 \pi)^3} \delta(\epsilon_\bk - \epsilon_F) = \frac{\epsilon_F}{4 \pi^2 v_F^2} \int_{-\pi/d}^{\pi/d} d k_z \Theta(\epsilon_F - |m(k_z)|)$
is the density of states at Fermi energy. We have also defined the average of $m(k_z)$ over the Fermi surface as
\beq
\label{eq:8}
\langle m \rangle = \frac{1}{g(\epsilon_F)} \int \frac{d^3 k}{(2 \pi)^3} m(k_z) \delta(\epsilon_\bk - \epsilon_F). 
\eeq

The impurity averaged response function, analytically continued to real frequency as $\Pi(\bq, \Omega) = \Pi(\bq, i \Omega \rightarrow \Omega + i \eta)$,
is given, in the self-consistent non-crossing approximation, by the sum of ladder diagrams, which gives
\beq
\label{eq:9}
\Pi(\bq, \Omega) = \Pi^I(\bq, \Omega) + \Pi^{II}(\bq, \Omega), 
\eeq
where 
\beq
\label{eq:10}
\Pi^I(\bq, \Omega) = 2 e^2 v_F \Omega \int_{-\infty}^{\infty} \frac{d \epsilon}{2 \pi i} \frac{d n_F(\epsilon)}{d \epsilon} P_{0 x}(\bq, \epsilon - i \eta, \epsilon + \Omega + i \eta), 
\eeq
and 
\beq
\label{eq:11}
\Pi^{II}(\bq, \Omega) = 4 i  e^2 v_F \int_{-\infty}^{\infty} \frac{d \epsilon}{2 \pi i} n_F(\epsilon) \textrm{Im} P_{0 x}(\bq, \epsilon + i \eta, \epsilon + \Omega + i \eta). 
\eeq 
The 4$\times$4 matrix $P$, whose $0 x$ component we are interested in, is given by $P(\bq, - i \eta, \Omega + i \eta) = \gamma^{-2} I^{RA}(\bq, \Omega) D(\bq, \Omega)$, 
$P(\bq, \epsilon + i \eta, \epsilon + \Omega + i \eta) = I^{RR}(\epsilon, \bq, \Omega)$, 
where $D = (1 - I^{RA})^{-1}$ is the diffusion propagator and $I^{RA}_{\alpha \beta}(\bq, \Omega) = \frac{\gamma^2}{2} \tau^{\alpha}_{\sigma_2 \sigma_1} \tau^{\beta}_{\sigma_3 \sigma_4} \int \frac{d^3 k}{(2 \pi)^3} G^R_{\sigma_1 \sigma_3}(\bk + \bq, \Omega) G^A_{\sigma_4 \sigma_2}(\bk , 0)$, 
$I^{RR}_{\alpha \beta}(\epsilon,\bq, \Omega) = \frac{1}{2} \tau^{\alpha}_{\sigma_2 \sigma_1} \tau^{\beta}_{\sigma_3 \sigma_4} 
\int \frac{d^3 k}{(2 \pi)^3} G^R_{\sigma_1 \sigma_3}(\bk + \bq,\epsilon + \Omega) G^R_{\sigma_4 \sigma_2}(\bk, \epsilon)$. 
The physical meaning of the two distinct contributions to the response function $\Pi^{I,II}(\bq, \Omega)$ is clear from Eq.~\eqref{eq:11}. 
$\Pi^{I}(\bq, \Omega)$ describes the non-equilibrium part of the response that happens at the Fermi surface. This response is 
diffusive when $\Omega \tau \ll 1$ and ballistic in the opposite limit. We will discuss this in more detail below. 
In contrast, $\Pi^{II}(\bq, \Omega)$ is an equilibrium, nondissipative contribution to the overall response, to which all states below 
the Fermi energy contribute~\cite{Streda}. 

We will start by evaluating the nonequilibrium part of the response function, $\Pi^I(\bq, \Omega)$. 
Computing the matrix elements $I^{RA}_{\alpha \beta}(\bq, \Omega)$ is easily done in the standard way, assuming $\epsilon_F  \tau \gg 1$. 
One obtains
\beq
\label{eq:14}
\Pi^I(\bq, \Omega) = i e^2 v_F \Omega \tau g(\epsilon_F) [I^{RA}_{00} D_{0x} + I^{RA}_{0x} D_{xx} + I^{RA}_{0z} D_{zx}], 
\eeq
where we have taken into account that $D_{yx} = 0$ by symmetry. 
The relevant matrix elements of the diffusion propagator can be found analytically to first order in $\bq$. One obtains
\beq
\label{eq:15}
\Pi^I(\bq, \Omega) = i e^2 v_F \Omega \tau g(\epsilon_F) \frac{I^{RA}_{0x} (1 - I^{RA}_{zz}) + I^{RA}_{0z} I^{RA}_{zx}}{\Gamma (1- I^{RA}_{xx})}, 
\eeq
where $\Gamma(\bq, \Omega) = (1 - I^{RA}_{00})(1 - I^{RA}_{zz}) - I^{RA}_{0z} I^{RA}_{z0}$ is the determinant  of the $0z$ block of the diffusion propagator, which corresponds to the diffusion of the charge density, 
a conserved quantity (this block decouples from the rest of the diffuson when $\bq \rightarrow 0$). 
This means, in particular, that $\Gamma$ must satisfy an exact Ward identity $\Gamma(0,0) = 0$.

Explicitly, the relevant matrix elements of $I^{RA}(\bq, \Omega)$ to first order in $\bq$ are given by
\beqa
\label{eq:17}
&&I^{RA}_{00} =\left\langle \frac{\tau_+/\tau}{1 - i \Omega \tau_+} \right\rangle, \,\, 
I^{RA}_{0x} = \frac{i v_F q}{2 \epsilon_F} \left\langle \frac{m }{\epsilon_F} \frac{\tau_+/\tau}{1 - i\Omega \tau_+} \right\rangle, \nonumber \\
&&I^{RA}_{0z} = I^{RA}_{z0} = \left\langle \frac{m}{\epsilon_F} \frac{\tau_+/\tau}{1 - i \Omega \tau_+}\right\rangle, \nonumber \\
&&I^{RA}_{zx} = \frac{i v_F q}{4 \epsilon_F} \left\langle \left(1+ \frac{m^2}{\epsilon_F^2}\right) \frac{\tau_+/\tau}{1 - i\Omega \tau_+} \right\rangle, \nonumber \\
&&I^{RA}_{zz} =\left\langle \frac{m^2}{\epsilon_F^2} \frac{\tau_+/\tau}{1 - i \Omega \tau_+} \right\rangle, \nonumber \\
&&I^{RA}_{xx} = \frac{1}{2} \left\langle \left(1- \frac{m^2}{\epsilon_F^2}\right) \frac{\tau_+/\tau}{1 - i\Omega \tau_+} \right\rangle,
\eeqa
where the average over the Fermi surface is defined in the same way as in Eq.~\eqref{eq:8}. 
The charge conservation Ward identity then takes the following explicit form
\beq
\label{eq:18}
\Gamma(0,0) = \left(1 - \left\langle \frac{\tau_+}{\tau} \right\rangle \right)  \left(1 - \left\langle \frac{m^2 \tau_+}{\epsilon_F^2 \tau} \right\rangle \right) - 
\left\langle  \frac{m \tau_+}{\epsilon_F \tau} \right\rangle^2 = 0. 
\eeq
The correctness of Eq.~\eqref{eq:18} may be easily checked using Eq.~\eqref{eq:6} and expanding in Taylor series in $m/\epsilon_F$. 
\begin{figure}[t]
\subfigure[]{
   \label{fig:1a}
  \includegraphics[width=7cm]{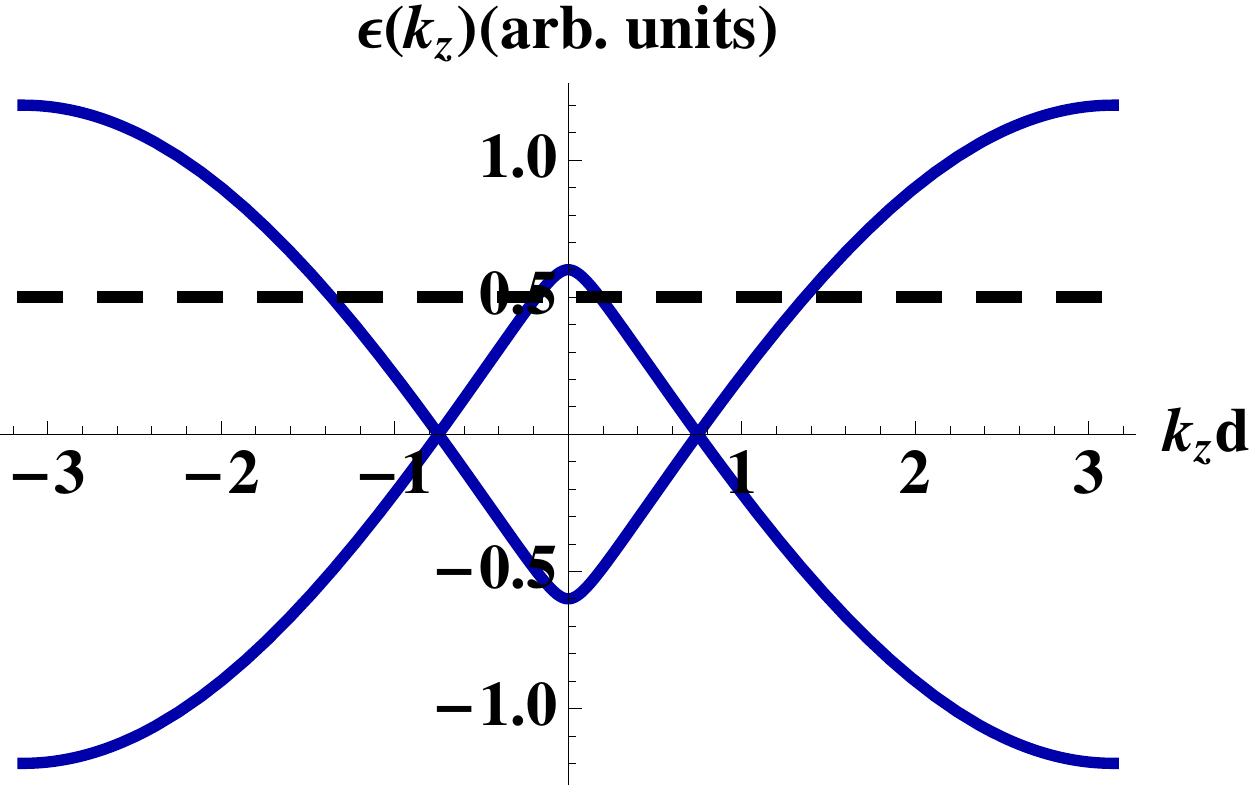}}
\subfigure[]{
  \label{fig:1b}
   \includegraphics[width=7cm]{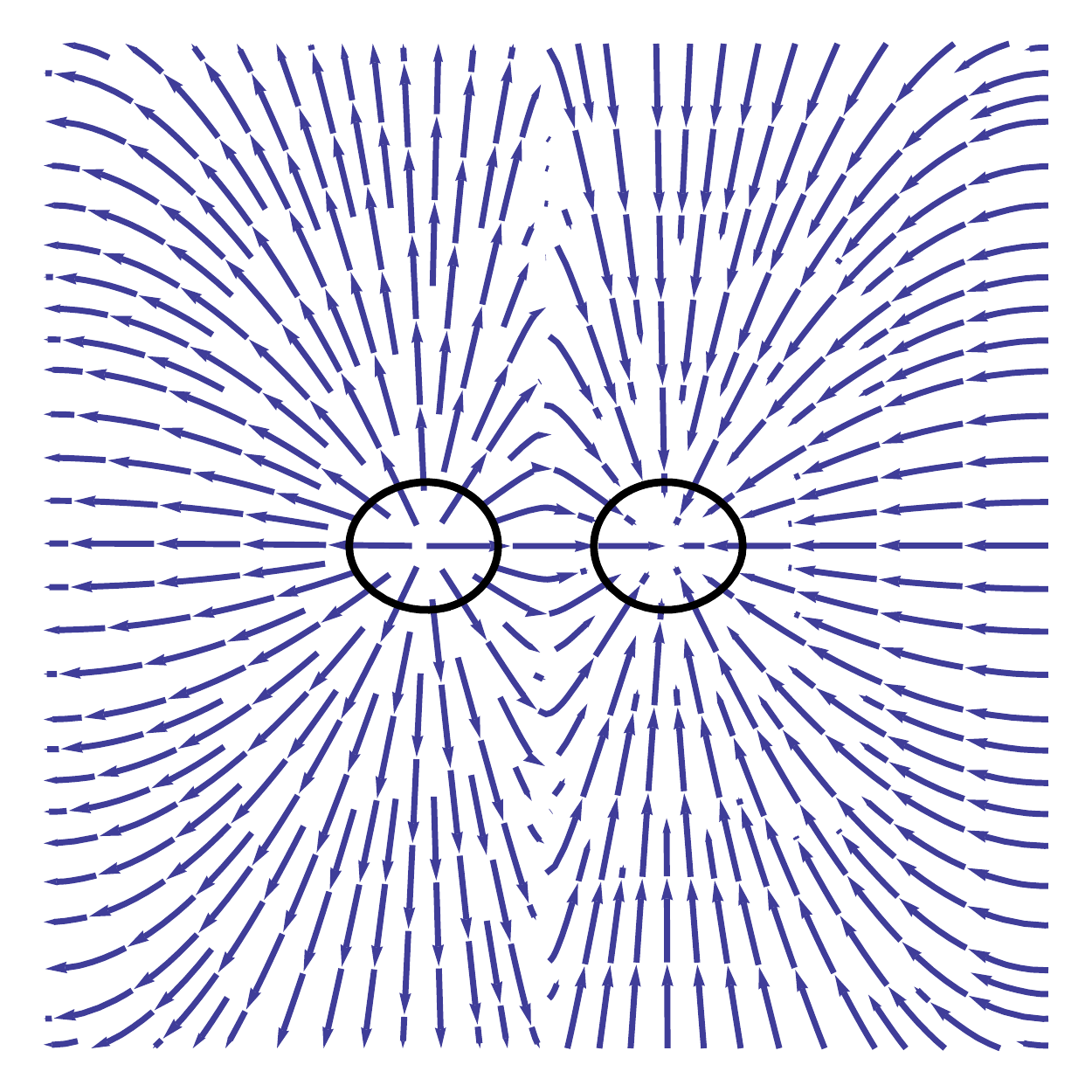}}
  \caption{(Color online). (a) Plot of the band edges along the $z$-direction in momentum space for the two bands that touch at the Weyl nodes, using specific 
  expression for $\Delta(k_z)$ from the multilayer model of Ref.~\cite{Burkov11}.  (b) Field lines of the Berry curvature 
  in the $k_y = 0$ plane, for the same band structure as in (a). Corresponding Fermi surface section is shown by the two contours, enclosing the Weyl nodes.}
    \label{fig:1}
\end{figure} 
Expanding $\Gamma(0,\Omega)$ to first order in $\Omega$ and taking the limit $\Omega \rightarrow 0$ at fixed $\tau$, which corresponds to the diffusive 
limit, we finally obtain
\beq
\label{eq:19}
\Pi_{dif}^I(\bq, 0) =  - \frac{i q \,e^2 v_F^2 g(\epsilon_F)}{2 \epsilon_F}\left\langle \frac{m \tau_+}{\epsilon_F \tau} \right\rangle F[m], 
\eeq
where
\beq
\label{eq:20}
F[m] = \frac{1 + \frac{1}{2} \left\langle \left(1 - \frac{m^2}{\epsilon_F^2} \right) \frac{\tau_+}{\tau}\right\rangle}
{1 - \frac{1}{2} \left\langle \left(1 - \frac{m^2}{\epsilon_F^2} \right) \frac{\tau_+}{\tau}\right\rangle}
\left[\left.\frac{\partial \Gamma(0,\Omega)}{\partial (\Omega \tau)}\right|_{\Omega = 0}\right]^{-1}. 
\eeq 
The explicit form of the functional $F[m]$ is in fact not that important for our purposes, except for the evenness property,
easily seen from Eq.~\eqref{eq:20}: $F[m] = F[-m]$. 
As a consequence, $\Pi_{dif}^I$ is an odd functional of $m$, which will play an important role below.
It is important to note that the charge conservation, whose mathematical consequence is the presence of the diffusion 
pole in $\Pi^I(\bq, \Omega)$, is crucial in obtaining a nonzero result in the diffusive limit in Eq.~\eqref{eq:19}. 
The analogous quantity in the calculation of the spin Hall conductivity, for example, would vanish in the diffusive limit~\cite{SHE}.

It is also of interest to examine the ballistic limit of $\Pi^I$, which corresponds to the case of a clean Weyl metal. In this case 
we send both $\Omega$ and $1/\tau$ to zero in such a way that $\Omega \tau \rightarrow \infty$. 
In this case we obtain
\beq
\label{eq:21}
\Pi^I_{bal}(\bq,0) = - \frac{i q \, e^2 v_F^2  g(\epsilon_F)}{2 \epsilon_F} \left\langle \frac{m}{\epsilon_F} \right\rangle, 
\eeq
which agrees with the clean Weyl metal result, obtained by us before~\cite{Burkov14,Pesin14}.     
As seen from Eqs.~\eqref{eq:19} and~\eqref{eq:21}, the difference between $\Pi^I_{dif}$ and $\Pi^I_{bal}$ is 
only quantitative. In the AHE literature, this difference is said to arise from the 
so-called side-jump processes~\cite{Molenkamp06,Nagaosa06,MacDonald07,Sinova10,Niu11}. 
\begin{figure}[t]
  \includegraphics[width=8cm]{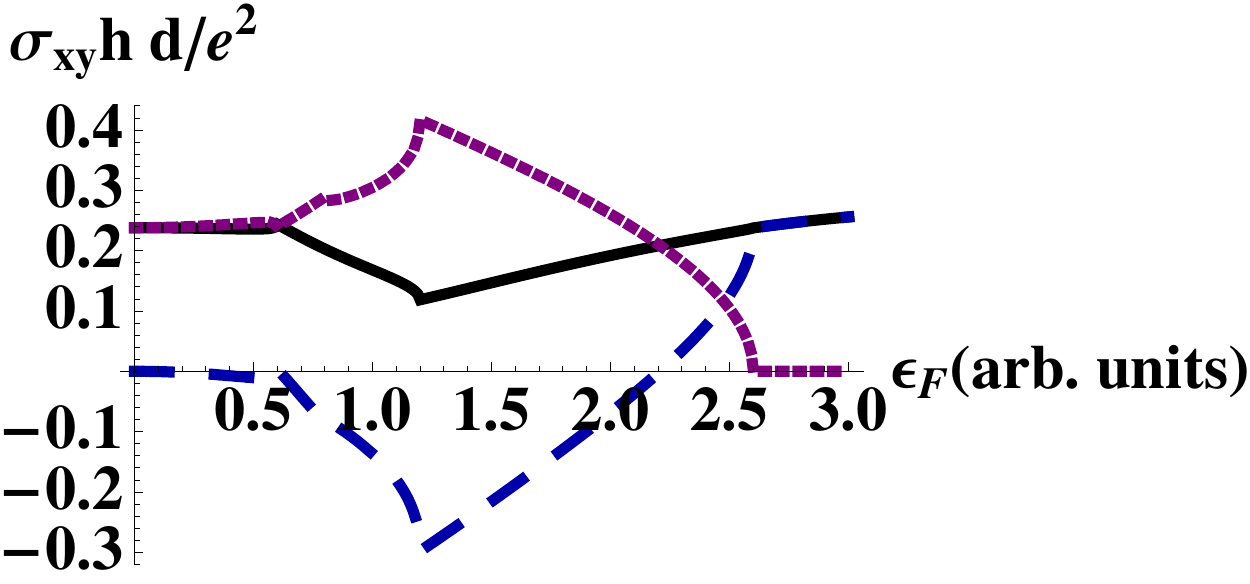}
  \caption{(Color online). Plot of the total anomalous Hall conductivity (solid line), $\sigma^I_{xy}$ (dashed line) and $\sigma^{II}_{xy}$ (dotted line) versus the Fermi energy for the same system as in Fig.~\ref{fig:1}.
  The plateau-like feature in $\sigma_{xy}$ correlates with the range of the Fermi energies, for which the Fermi surface consists of two separate sheets, each enclosing 
  a single Weyl node. The energy units are the same as in Fig.~\ref{fig:1}.}
  \label{fig:2}
\end{figure}
The final step of the calculation is to evaluate the equilibrium part of the response function, $\Pi^{II}(\bq, \Omega)$. 
In the limit $\epsilon_F \tau \gg 1$ one finds that this part of the response function is unaffected by the impurity scattering and is 
given by
\beqa
\label{eq:22}
\Pi^{II}(\bq, \Omega)&=&e^2 v_F \int \frac{d^3 k}{(2 \pi)^3} \langle z^{s'}_\bk | z^s_{\bk + \bq} \rangle \langle z^s_{\bk + \bq} | \tau^x | z^{s'}_\bk \rangle \nonumber \\
&\times&\frac{n_F(\xi^s_{\bk + \bq} - \Omega) - n_F(\xi^{s'}_\bk)}{\Omega - \xi^s_{\bk + \bq} + \xi^{s'}_\bk}, 
\eeqa
where summation over the band indices $s, s'$ is again implicit. 
Evaluating Eq.~\eqref{eq:22} in the small $\Omega$ and $q$ limit gives 
\beqa
\label{eq:23}
\Pi^{II}(\bq, 0)&=&\frac{-i q \, e^2}{8 \pi^2} \int_{-\pi/d}^{\pi/d} d k_z \textrm{sign}[m(k_z)] \nonumber \\
&\times&\left\{1 - \Theta[\epsilon_F  - |m(k_z)|]\right\}.
\eeqa
The first term in Eq.~\eqref{eq:23} arises from the completely filled bands, while the second is the contribution of the incompletely filled bands. 

We can now finally evaluate the anomalous Hall conductivity. We will focus on the diffusive limit results, as ballistic limit is qualitatively similar. 
At this point we will also explicitly include the contribution of both $t = \pm$ pairs of bands, which simply amounts to restoring the index $t$ in $m_t$, and 
summing over $t$. 
Using Eq.~\eqref{eq:3} and remembering that $A_0 \rightarrow i A_0$ upon Wick rotation to the real time, we obtain
\beq
\label{eq:24}
\sigma^I_{xy} = \frac{e^2 v_F^2}{2 \epsilon_F} \sum_t g_t(\epsilon_F) \left\langle \frac{m_t \tau_{+ t}}{\epsilon_F \tau_t} \right\rangle F[m_t], 
\eeq
and 
\beqa
\label{eq:25}
\sigma^{II}_{xy}&=&\frac{e^2}{8 \pi^2} \sum_t \int_{-\pi/d}^{\pi/d} d k_z \textrm{sign}[m_t(k_z)] \nonumber \\
&\times&\left\{1 - \Theta[\epsilon_F  - |m_t(k_z)|]\right\}.
\eeqa
Since $m_+(k_z)$ is positive throughout the first BZ, while $m_-(k_z)$ changes sign at the Weyl nodes, the first term 
in Eq.~\eqref{eq:25}, which comes from completely filled bands, gives a universal (almost) quantized contribution
\beq
\label{eq:26}
\sigma^{quant}_{xy} = \frac{e^2 {\cal K}}{4 \pi^2}, 
\eeq
where ${\cal K}$ is the distance between the Weyl nodes. This equation also describes the cases when the Weyl nodes are absent, 
in which case $\sigma^{quant}_{xy}$ is truly quantized since ${\cal K} = 0, G$, where $G= 2\pi /d$ is a reciprocal lattice vector. 

We are now ready to state our main result. This comes from examining the remaining, non-quantized parts of $\sigma_{xy}$. 
Suppose we have a situation when the Weyl nodes are present and $\epsilon_F$, while not zero, is not too far from it, 
as shown in Fig.~\ref{fig:1}.
Recall that at the location of the Weyl nodes $m_-(k_z) = b - \Delta(k_z)$ changes sign. 
This implies that, as long as ${\cal K} \left.\frac{d \Delta}{d k_z}\right|_{k_z = k_{z0}} \gg \epsilon_F$, 
where $k_{z0}$ is the location of a given Weyl node, the average of any odd function of $m_-(k_z)$ over the Fermi surface will vanish. 
This means that in such a situation, which we call {\em Weyl metal}, all contributions to the anomalous Hall conductivity, 
associated with incompletely filled bands, will vanish and $\sigma_{xy}$ attains a universal value, characteristic 
of Weyl semimetal $\sigma_{xy} = \sigma_{xy}^{quant}$, where $\sigma_{xy}^{quant}$ is given by Eq.~\eqref{eq:26}~\cite{footnote2}. 
This is illustrated in Fig.~\ref{fig:2}.
Note that the linear dispersion sufficiently close to Weyl nodes is a topological property, 
in the sense that it follows directly and exclusively  from the existence of a nonzero topological charge by the so-called Atiyah-Bott-Shapiro construction~\cite{Horava}.  

To understand this result physically, it is helpful to recall that the Weyl nodes are monopole sources of the Berry curvature $\bOmega_\bk$. 
In a clean metal, the anomalous Hall conductivity $\sigma_{xy}$ is given by the integral of the $z$-component of the Berry curvature over all 
occupied states $\sigma_{xy} = e^2 \int \frac{d^3 k}{(2 \pi)^3} n_F(\epsilon_\bk) \Omega^z_\bk$.
However, as clear from Fig.~\ref{fig:1}, when the Fermi surface breaks up into disconnected sheets, enclosing individual nodes, 
the contribution of the states, enclosed by the Fermi surface, to this integral will always be very small, vanishing exactly in the limit when the band dispersion away from the nodes may be taken to be exactly linear.
An obvious analogy here is with the electric field of a dipole. A pair of Weyl nodes is like a dipole of two topological charges. Its field has a well-defined and nonzero on average 
$z$-component at large distances from the dipole. At short distances, however, the field is that of individual charges, which winds around the location 
of each charge and thus any particular component of it averages to zero. 

In conclusion, we have demonstrated that the AHE in Weyl metals has a purely intrinsic origin and can be associated entirely with the Weyl nodes, just 
as in the case of a Weyl semimetal, when the Fermi energy coincides with the nodes and the Fermi surface is absent. 
This is in contrast to an ordinary FM metal, in which the anomalous Hall conductivity always has both a significant Fermi surface contribution and 
an extrinsic contribution.

\begin{acknowledgments}
Financial support was provided by Natural Sciences and Engineering Research Council of Canada. 
\end{acknowledgments}


\begin{thebibliography}{99}
\bibitem{footnote1} Note that Eq.~\eqref{eq:1} has an explicit particle-hole symmetry (which could be broken e.g. by adding diagonal terms, quadratic in $k_z$). 
The (at least approximate) existence of this symmetry is important, since otherwise the Weyl semimetal phase, from which the Weyl metal is 
obtained by doping, will not exist (there will always be a finite volume Fermi surface, in addition to the Weyl nodes), and our analysis will then not apply.
\bibitem{footnote2} While we have explicitly ignored Coulomb interactions, we do not expect this to be a problem, since our results originate
from symmetry properties of the electronic structure in the vicinity of the Weyl nodes.
\end{thebibliography}
\end{document}